\newcommand\Mpc{$h^{-1}$Mpc}
\newcommand\etal{{\it et al.\ }}
\newcommand\kmsec{${\rm km~sec^{-1}}$}
\newcommand\sig{\sigma_{12}}

\newcommand\s{{\bf s}}

\newcommand\R{{\cal R}}
\newcommand\F{{\cal F}}

\documentstyle[12pt,aaspp4,epsf]{article}
\begin{document}

\title{The Pairwise Velocity Distribution of Galaxies in the Las
Campanas Redshift Survey}

\author{Stephen D. Landy}

\affil{Department of Physics, College of William and Mary,
     Williamsburg, VA 23187-8795
        landy@physics.wm.edu}

\author{Alexander S. Szalay}

\affil{Dept. of Physics and Astronomy, Johns Hopkins
University, Baltimore, MD 21218
szalay@pha.jhu.edu}

\and

\author{Thomas J. Broadhurst}

\affil{Astronomy Dept., 601 Campbell Hall, UC Berkeley, Berkeley, CA 94720
tjb@astron.berkeley.edu}

\begin{abstract}

We present a novel measurement of the pairwise peculiar velocity
distribution function of galaxies on scales $r<3200$ \kmsec~in the
Las Campanas Redshift Survey. The distribution is well described by a
scale-independent exponential with a width $\tau$ where $\sig
=\sqrt{2}\tau=363$ \kmsec.  The signal is very stable. Results from
the Northern and Southern sets of slices agree within $\pm13$
\kmsec~and the fluctuations among the six individual survey slices
vary as $\pm44$ \kmsec. The distribution was determined using a
Fourier-space deconvolution of the redshift-space distortions in the
correlation function. This technique is insensitive to the effect of
rich clusters in the survey and recovers the entire distribution
function rather than just its second moment. Taken together with the
large effective volume of the survey $6.0\times10^6$~\Mpc$^3$, we
believe this to be a definitive result for $r$-band selected galaxies
with absolute magnitudes $-18.5 <{\rm M_r}<-22.5$ and $z<0.2$.

\end{abstract}


\section{Introduction}

The pairwise peculiar velocity dispersion of galaxies is an important
measure of the structure and clustering of the universe on
large-scales and is used as a test of cosmological models. In a
seminal paper, Davis and Peebles (1983) using redshift information
from the first Center for Astrophysics Redshift Survey (CfA1) measured
a value of $\sig= 340\pm40$~\kmsec. Subsequent numerical work
predicted a much larger value of approximately 1000~\kmsec~for a
standard $\Omega h=0.5$ Cold Dark Matter (CDM) model (Davis
\etal 1985).

Additional redshift surveys have given somewhat discrepant
results. The IRAS 1.2 Jy survey (Fisher \etal 1994) was in good
agreement with a value of $\sig= 317^{+40}_{-49}$~\kmsec~while Marzke
\etal (1995) using the second Center for Astrophysics Redshift Survey
(CfA2) together with the Southern Sky Redshift Survey (SSRS) found
$\sig= 540\pm180$~\kmsec. Similar analysis applied to the Las Campanas
Redshift Survey has found $\sig=452\pm60$~\kmsec (Lin \etal
1995). Guzzo \etal (1997) find $\sig= 345^{+95}_{-65}$~\kmsec~for
late-type galaxies using the Pisces-Perseus Redshift Survey and report
this to be a fair estimate for field galaxies.

Numerical work and re-analysis of existing surveys have shown the
sensitivity of the measurement of the velocity dispersion to the
idiosyncracies of the data using standard techniques. Mo \etal (1993)
found that the estimated dispersion is extremely sensitive to the
presence of rich clusters in a sample with values ranging from $\sig=$
300 to 1000 \kmsec~for subsets of the same data. Zurek \etal (1994)
using high resolution CDM simulations also found large variations in
the value of the dispersion on CfA1 size scales. Moreover, Marzke
\etal (1995) found that by excluding the rich clusters from their
survey, Abell clusters with R$\ge 1$, the measured velocity dispersion
dropped to $\sig=295\pm99$~\kmsec.

From a purely analytical perspective, redshift surveys present serious
challenges to the measurement of the velocity dispersion. A
fundamental difficulty is simply that a direct measurement of the
velocity dispersion requires galaxy positions and peculiar velocities
whereas it is only galaxy redshifts which are available. Therefore an
indirect approach has been taken which exploits the anisotropies
created by the peculiar velocities in the redshift space correlation
function (see Peebles 1980). In this method, the redshift space
correlation function is represented in two dimensions, where the axes
correspond to the directions parallel and perpendicular to the
line-of-sight separation of a galaxy pair. The resultant correlation
function is anisotropic since the peculiar velocities distort the
correlation function principally along the line-of-sight.  These
anisotropies are then used to estimate the peculiar velocity
dispersion of galaxies on various scales.  An inherent problem is that
the velocity distribution function distorts the correlation function
by way of a convolution. The galaxy peculiar velocity distribution
function $f(w_3|r)$ is usually modelled as an exponential along one
dimension $(w_3)$ with

\begin{equation}
f(w_3|r)={1\over{\sqrt{2}\sig}}
\exp{\left(-{\sqrt{2}|w_3-v_{12}|\over\sig}\right)}.
\end{equation}

Thus, the exponential is characterized by its width, $\tau =
\sig/\sqrt{2}$, corresponding to the thermal velocity dispersion of
galaxy pairs (see Peebles 1980 for the physical motivation behind this
parameterization). In practice, this is determined from computing the
anisotropic second moment at a number of separations between $r=1$ and
$r=5$ Mpc, {\it e.g.} Davis and Peebles (1983), Davis \etal (1985),
Fisher \etal (1994), Lin \etal (1995).  The infall $v_{12}$ is
difficult to model, since in general it is scale dependent, and
correlated with $\sigma$, but it is negligible for small scales.
The difference between the dispersions in the line-of-sight and
transverse direction is used as an estimator for $\sig$.  This
quantity is particularly sensitive to the presence of highly
virialized rich clusters, which have a pronounced effect on the
estimate of the second moment.

Our goal was to develop a robust method to measure the distribution
function with a minimum of assumptions as to its form and with a
decreased sensitivity to rich clusters. This was accomplished by
considering the Fourier transform of the two dimensional redshift
space correlation function in the non-linear regime. The convolution
of the distribution function with the correlation function becomes a
multiplication in Fourier space. With a good signal-to-noise data, as
is the case for the LCRS, a direct deconvolution is possible. This
approach turns out to be straight-forward and the form of the
distribution function can be directly recovered. A further advantage
of working in Fourier space is that the tail of the distribution
affects only a few pixels at small wavenumbers, resulting in a greatly
reduced sensitivity to the presence of rich clusters. This technique
will be developed and applied in the following {\it Letter}. The data
will be described in Section 2, the method in Section 3, and the
results in Section 4. The discussion and conclusion are in Section 5.

\section{Data}

The data is that of the Las Campanas Redshift Survey which has been
described in Shectman \etal (1996). In short, the full data set
consists of over 26,000 galaxies in six slices each approximately
1.5\arcdeg\ thick in declination by 80\arcdeg\ wide in right ascension
and is an r-band selected sample with nominal apparent magnitude
limits of $15.0 <{\rm m_r}<17.7$ and $z<0.2$. In each hemisphere the
slices are centered on the same right ascension while being offset in
declination by 3 or 6 degrees. The three slices in the South galactic
hemisphere are located at -39\arcdeg,- 42\arcdeg, and -45\arcdeg\ Dec
centered on $0^h45^m$ RA, and the three in the North at
-3\arcdeg,-6\arcdeg, and - 12\arcdeg\ Dec centered on $12^h45^m$ RA.

In a manner similar to Landy \etal (1996), the data set was culled to
redshifts between 10,000 \kmsec~ and 45,000 \kmsec~with absolute
magnitude limits of $-18.5 <{\rm M_r}<- 22.5$, leaving 19,306
galaxies. Values of $q_o=0.5$ and $H_o = 100 h~$\kmsec~Mpc$^{-1}$ were
used to convert to co-moving distances.  The slices were analysed
individually and as combined sets of three for the North and South
hemispheres.  Throughout this analysis the galaxies are point
weighted.  Issues concerning fiber optic separation limitations and
undersampling will be discussed in Section 4.

\section{Method}

\subsection{Correlation Function}

Following the usual treatment of Davis and Peebles (1983), we compute
the redshift space galaxy correlation function in terms of two
orthogonal coordinates, one perpendicular ($r_p$) and the other
parallel ($\pi$) to the line-of-sight. These form a vector $\s_{ij}$
joining each pair of galaxies $(i,j)$. The galaxies have position
vectors $\s_i$ and $\s_j$. However, we deviate somewhat from the usual
coordinates, in that we define the vector which bisects the angle
between $\s_i$ and $\s_j$ as the direction of the line-of-sight rather
than the direction of the center of mass of the two galaxies as has
been common in earlier work (see Peebles 1979; Davis \etal 1978; Davis
\& Peebles 1983; Fisher \etal 1994). The advantages of this coordinate
system are described in detail in Szalay, Matsubara, \& Landy
(1997). Such a parametrization has been used in a different context by
Hamilton (1992). The two coordinate systems coincide in the
plane-parallel case, i.e. when the angle between the two line of sight
vectors is small, which is the case for this analysis.

The probability that any two galaxies are separated by $r_p$ and $\pi$
is given by
\begin{equation}
        \delta P=[1+\xi(r_p,\pi)] 2\Pi r_p \delta r_p\delta\pi
\end{equation}
where $\xi(r_p,\pi)$ is the anisotropic, two dimensional
redshift space correlation function. To estimate
$\xi(r_p,\pi)$ from the data we use

\begin{equation}
        \xi(r_p,\pi) = {{DD(r_p,\pi)- 2DR(r_p,\pi)+RR(r_p,\pi)}
        \over RR(r_p,\pi)}
\end{equation}
where $DD$, $DR$, and $RR$ are the normalized distribution of
data-data, data-random, and random-random pairs of galaxies in the
same geometry as the survey (see Landy \& Szalay 1993).

\subsection{Fourier Analysis}

Up to this point, our approach has been very similar to earlier
work. Plate 1a shows the mean of $\xi(r_p,\pi)$ for the Northern and
Southern sets of slices. This function has been reflected about each
axis and the distortion of the contours along the $\pi$ axis due to
the pairwise galaxy peculiar velocity distribution is easily seen. The
correlation function was estimated using bins of width 25 \kmsec in
both the $r_p$ and $\pi$ directions and has been truncated at 3200
\kmsec in order to isolate the distortions in the non-linear regime.

Next, this function was multiplied with a 3200 \kmsec~wide 2D Hann
window centered on the origin, and a 2D Fourier transform
performed. The result is shown in Plate 1b. Since Fourier space is
dual to real space, functions which are broad in real space become
narrow in Fourier space and {\it vice-versa}. The width of the window
had no appreciable effect, as long as it remained above 1000
\kmsec.

Our motivation is to extract the pairwise peculiar velocity
distribution function. As is well known, convolution in real space
corresponds to multiplication in Fourier space. Let us consider taking
cuts along the $k_{r_p}$ and $k_\pi$ axes in the 2D power spectrum
denoted by $\R(k)$ and $\Pi(k)$, respectively. Due to the so-called
`slicing-projection theorem' the function $\R(k)$ is simply the
Fourier transform of the projection of the correlations onto the $r_p$
axis. This projection, since we collapse all information along the
line-of-sight, is independent of the distribution function or of any
redshift-space distortions, since for most of our pairs, at such a
small separation, the two lines of sight are close to parallel.  On
the other hand, the function $\Pi(k)$ is the Fourier transform of the
projection of the correlations onto the $\pi$ axis. This function can
be modelled as the Fourier transform of the convolution of the
underlying correlation function $\xi(r)$ with the distribution
function. This shows, that even though we used only a small fraction
of the data in the Fourier-domain, via the cuts, we are still using
information about all the pairs.

By Occam's razor, let us take the distribution function $f(w_3|r)$ in
Eq (1) to be independent of $r$, and neglect the effects of infall.
Since our weighting scheme is such, that we are mostly sensitive to
the value of $\sigma$ at scales of $\approx 1$\Mpc, the infall
velocity due to its linear rise around the origin is indeed
negligible.  Let the Fourier transform of $f(w_3)$ be $\F(k)$.  Then
\begin{equation}
        \Pi(k)=\F(k) \R(k).
\end{equation}

Therefore, dividing the signal along the $k_{\pi}$ axis, $\Pi(k)$, by
the signal along the $k_{r_p}$ axis, $\R(k)$, returns $\F(k)$ - the
Fourier transform of the distribution function. This method is used
below with the survey data.

It should be noted that many other researchers have recognized the
advantages to working with the Fourier transform of the redshift space
correlation function in order to measure the redshift distortion
signal (see Kaiser 1987, Heavens \& Taylor 1995, Cole, Fisher, \&
Weinberg 1994,1995, and Zaroubi \& Hoffmann 1996). However, this
method differs in that it is designed to recover the distribution
function in the non-linear regime.

\section{Results}

The functions $\Pi(k)$ and $\R(k)$ for the Northern and Southern sets
of slices are shown in Figures 1a and 1b (all axes are shown with
units of frequency $f$).  The agreement between these two independent
samples is remarkable. The ratios of these functions for the two data
sets are shown in Figure 1c. This is the function $\F(k)$, which is
the Fourier transform of the distribution function. Here the data has
been truncated for $f>0.5$ ($\lambda<200$ \kmsec) to limit noise.

Since the results for the North and South are almost identical, the
average is graphed in Figures 2a and 2b along with the best fits to
the Fourier transforms of two functions commonly used to model the
distribution function, a Gaussian and an exponential. The Fourier
transform of a Gaussian is also a Gaussian while the Fourier transform
of a cuspy exponential is a Cauchy (or Lorentzian) distribution.  As
can be seen, the Lorentzian is an excellent fit to the data. Its
associated exponential has a width of $\sigma_{12}= 363$~\kmsec. This
exponential as well as $\F(k)$ taken directly from the data are shown
in Figure 2c. Here, transforming back to real space the agreement is
again remarkable.

In practice, the procedure was to take exponential distributions with
varying decay constants and window them with the same Hann window as
with the data. These functions were Fourier transformed and normalized
to a central value of one and then fit to the appropriate $\F(k)$
using a simple $\chi^2$ minimization routine. A summary of the fits to
various slices and sets is given in Table 1. The intrinsic variance of
the individual slices gives $\pm 44$~\kmsec~while the difference
between the Northern and Southern sets is only 10 \kmsec. Although
$\chi^2$ values are difficult to interpret for such a procedure, this
variance is consistent with an increase in the $\chi^2$ of a factor of
three. It should be noted that the two somewhat outlier slices 3 and
6, are also the most sparsely sampled of all slices in Las Campanas
survey.

\begin{deluxetable}{lrrrrcrrrrr}
\tablewidth{33pc}
\tablecaption{Pairwise Peculiar Velocity Dispersion Measurements (\kmsec)}
\tablehead{
        \colhead{Slice(s)}           &
        \colhead{$\sigma_{12}$}   &
        \colhead{$\Delta\sigma_{12}$}
}
\startdata
Mean North-South Sets & 363 & $\pm$13 \nl
North Set & 355 & \nl
South Set & 373 & \nl
-3        & 293 & \nl
-6        & 416 & \nl
-12       & 361 & \nl
Avg. of Northern Slices & 357 & \nl
-39       & 381 & \nl
-42       & 400 & \nl
-45       & 342 & \nl
Avg. of Southern Slices & 374 & \nl
\tablecomments{The Mean North-South signal was constructed
by taking the mean of the correlation functions for the North
and South sets of three slices. The variance is calculated from
considering the signals from the North and South sets
as independent measurements. A set reflects the correlation
function for the three respective slices considered
simultaneously and contains more galaxy pairs
since pairs of galaxies between slices are included as well.
The average is a simple average of three slices taken individually.
This is shown for completeness only since it is a not an optimal measure of
the signal.}
\enddata
\end{deluxetable}

\subsection{Idiosyncracies of the Las Campanas Data}

The Las Campanas Redshift Survey was constructed using a multi-fiber
optic spectrograph. Because of the size of the fibers, galaxies closer
than 55\arcsec~could not be observed simultaneously. Since each
spectroscopic field was observed only once, the survey contains an
intrinsic deficit of close pairs; on average six percent of the
galaxies originally chosen randomly to be observed were excluded. This
is of some concern to measurements of the pairwise velocity dispersion
as a bias may result which underestimates the velocity dispersion due
to the loss of these pairs.

As a check against this problem the data was identically analysed
except for the following variations and the results compared. First,
the data was analysed without regard to the missed galaxies. Second,
the partners of the missed galaxies were given a double weighting. And
third, the missed galaxies were assigned to the redshift of their
partner and a peculiar velocity of 250 \kmsec~drawn from an
exponential distribution was added to that redshift (500~\kmsec~ was
also used with similar results). These three procedures should bracket
the expected systematics as the first would be expected to
underestimate the velocity dispersion while the third should tend to
overestimate it. The second corresponds closely to the $\sigma{12}$
derived from the data. All of these results were well within the
$\pm44$\kmsec~ being reported as the intrinsic error. The results as
presented correspond to the third procedure.

\section{Discussion and Conclusions}

The velocity distribution was first characterized as an exponential by
Peebles (1976). Since that time many researchers have confirmed this
result (see Peebles 1979, Davis \& Peebles 1983, Bean \etal 1983,
Hale-Sutton \etal 1989, Fisher \etal 1994). However, it has been
assumed using well-motivated and general arguments that this
distribution should be a weak function of scale $\propto
r^{0.2}$. These results indicate the distribution function to be very
well represented by an exponential independent of scale in the
non-linear regime. Marzke \etal (1996) using the CfA2 redshift survey
also found a velocity distribution consistent with an exponential
distribution independent of scale.

In this paper we have developed a robust way to directly measure the
shape of the thermal velocity distribution of galaxies. Using a
Fourier deconvolution, we show that the distribution is extremely well
described by an exponential.  Both the shape and the width of
$\sigma_{12}=363\pm44$~\kmsec~are in excellent agreement with the
measurement of Davis and Peebles (1983). This measurement is also in
agreement with that of Guzzo \etal (1997) who find a value of
$\sig=345^{+95}_{-65}$~\kmsec~for late-type galaxies as a fair
estimate for field galaxies using the Pisces-Perseus Redshift Survey .

Our technique is a robust method. It is not pair weighted, even though
it uses the correlation function. Since we based our deconvolution
procedure on the 2D correlation function, we have effectively used a
weighting function proportional to $r_p^{-1}$. This emphasizes the
high signal-to-noise core of the correlation function. The presence of
rich clusters in the sample would superimpose a number of pairs with a
higher dispersion than the thermal dispersion of field galaxies.
Measurements of the dispersion via a second moment will be quite
sensitive even to a small number of pairs in the tail of the
distribution. Measuring the whole shape in Fourier space has the
advantage that the large number of bins in the tail are all compressed
into the innermost few resolution elements in Fourier space, therefore
they are naturally downweighted in our fitting procedure, which used
equal weights for every cell in $k$-space. It should be emphasized
that this method does not exclude the signal from the rich clusters
but rather incorporates them in a way which decreases their
contribution to the variance in a robust fashion.

Such a small value for the dispersion is quite interesting, especially
in comparison to the most recent theoretical estimation of the small
scale 1D velocity dispersion by Davis, Miller, and White (1997), who
predict a value of 600 \kmsec~ for an $\Omega_o=1$ Universe. This is
an `unfiltered' value, since it contains contributions from bulk flows
as well, thus the thermal component, that we are measuring can be
somewhat lower.

The tight agreement between average and the signal from the Northern
and Southern sets of slices is excellent, $\Delta\sig<$13 \kmsec,
considerably better than any previous measurement of this quantity.
The overall uncertainty is $\Delta\sig<$44 \kmsec~ when we consider
the individual slices and idiosyncracies of the data. This indicates
that $\sig=363$ \kmsec~ is a robust and probably a definitive measure
of the pairwise peculiar velocity dispersion in the local universe.

\section{Acknowledgements}

        The authors would like to acknowledge useful discussions
with Marc Davis.

\newpage

\epsfxsize=5.25in
\epsfbox{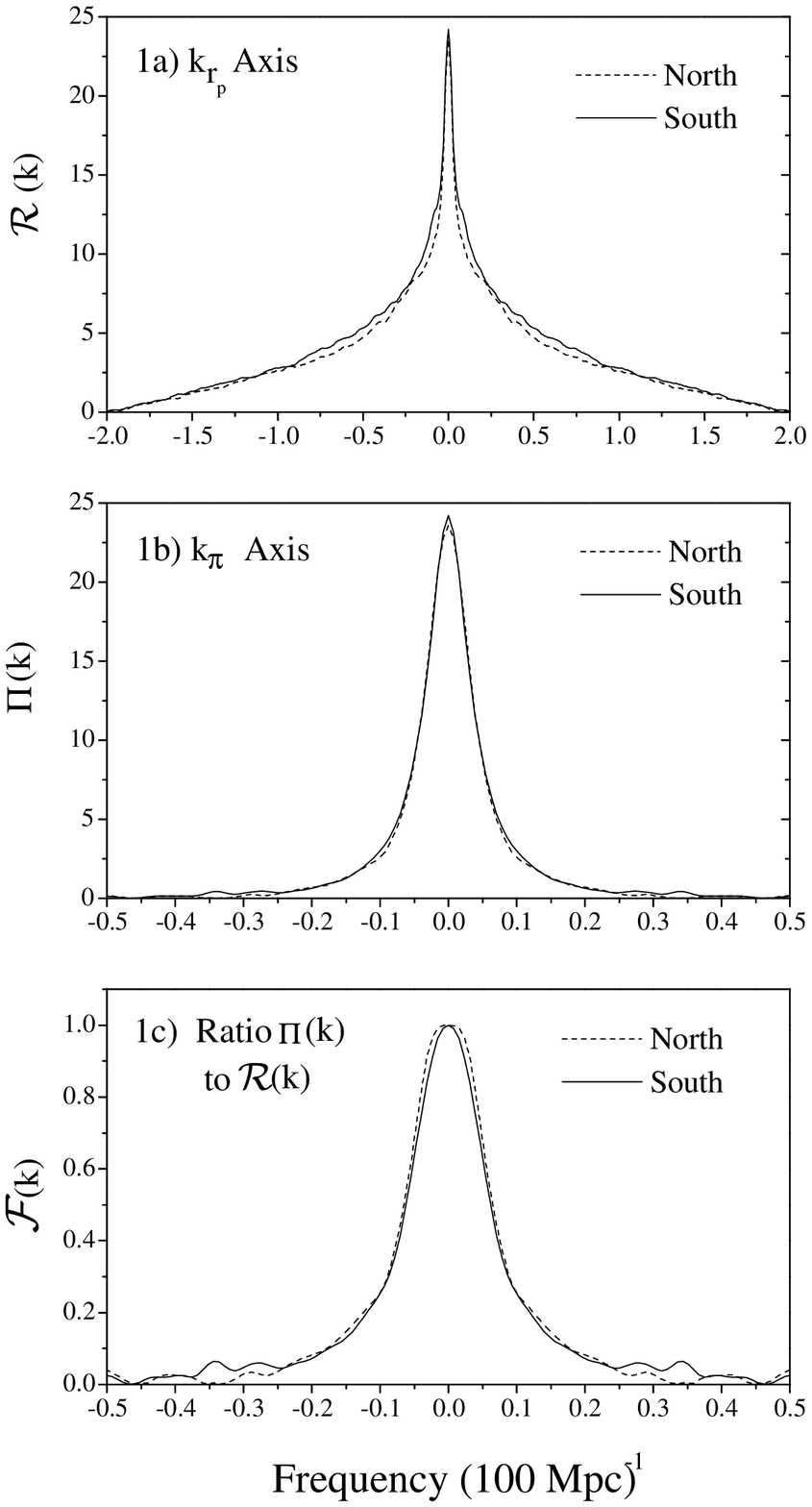}

Figure 1a shows the cuts in Fourier space along the
$k_{r_p}$ axis - $R(k)$. Figure 1b is the $k_{\pi}$ axis - $\Pi(k)$,
and Figure 1c is $F(k)$ - the ratio of $\Pi(k)$ to $R(k)$. This ratio
is the Fourier transform of the dispersion function in the non-linear
regime. Results for the combined sets of North and South slices are
shown. Notice the excellent agreement for these two independent data
sets.

\newpage

\epsfxsize=5.0in
\epsfbox{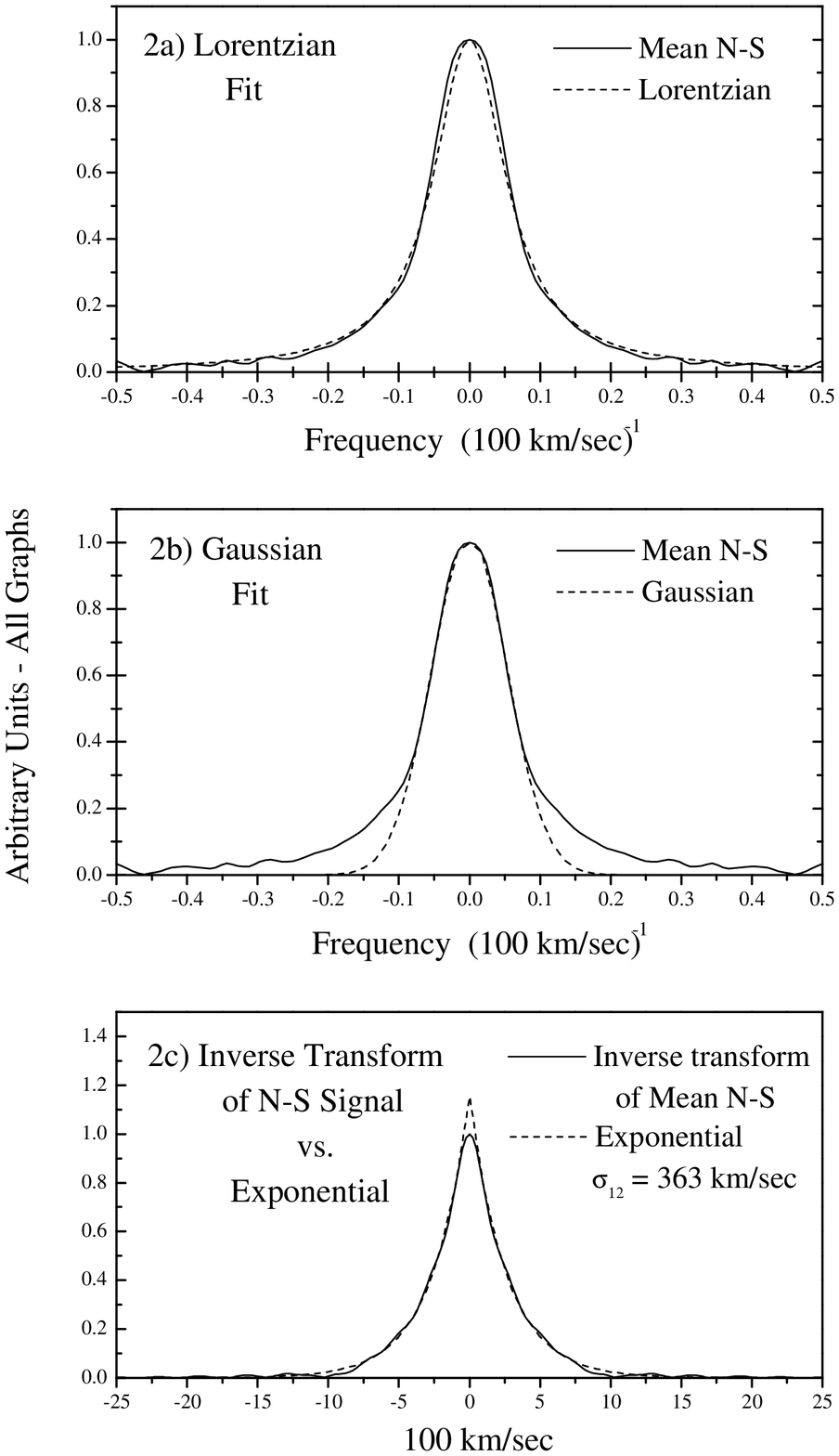}

Figure 2a is the best Lorentzian fit to the average of
the functions $F(k)$ for the North and South sets shown in Figure
1c. A Lorentzian is the Fourier transform of an exponential
distribution. A Gaussian fit is shown for comparison in Figure
2b. Figure 2c shows the Fourier transform of the functions in Figure
2a. The fit to the exponential is remarkable.

\newpage

\epsfxsize=4.5in
\epsfbox{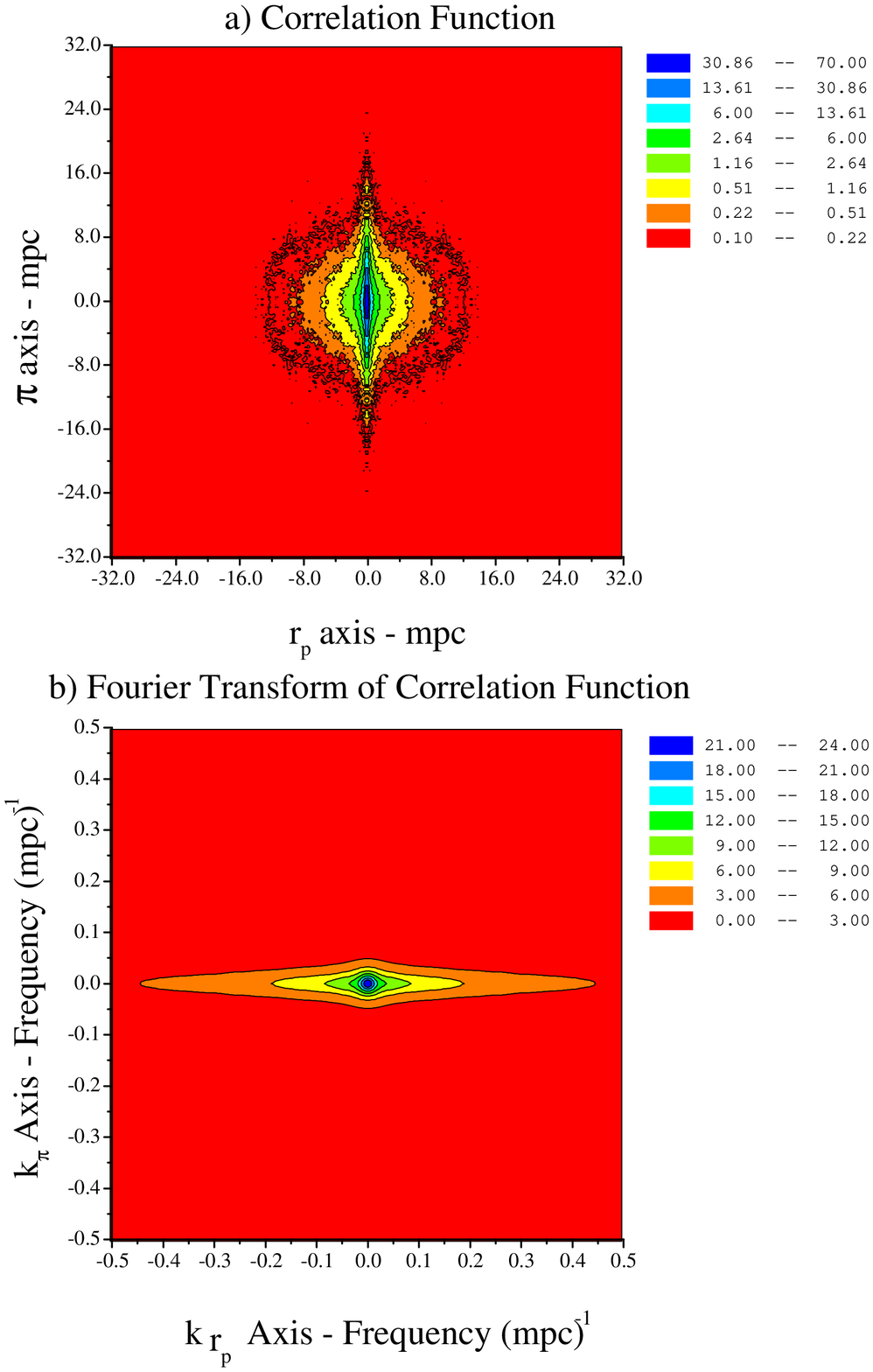}

Plate 1a) The correlation function $\xi(r_p,\pi)$ for
the Northern and Southern sets of slices which has been
reflected about each axis is shown. The distortion
of the contours along the $\pi$ axis due to the pairwise
galaxy peculiar velocity distribution is easily seen.

Plate 1b) The Fourier transform of the correlation function
multiplied with a 3200 \kmsec~wide Hann window is
shown. Since Fourier space is dual to real space, functions
which are broad in real space become narrow in Fourier space
and {\it vice-versa}. It is from the ratio of the value of this
function along the $k_{\pi}$ and $k_{r_p}$ axes that the dispersion
function is calculated.

\end{document}